\def\kF{k_{\text{F}}}

\def\dndmu{\partial n/\partial\mu}
\def\be{\begin{equation}}
\def\ee{\end{equation}}
\def\bea{\begin{eqnarray}}
\def\eea{\end{eqnarray}}
\def\bse{\begin{subequations}}
\def\ese{\end{subequations}}

\documentclass[prl,twocolumn,showpacs,preprintnumbers,amsmath,amssymb]{revtex4}
\usepackage{graphicx}
\usepackage{dcolumn}
\usepackage{bm}
\usepackage{color}
\begin{document}
\preprint{}
\title{Scaling Theory of a Compressibility-Driven Metal-Insulator Transition in a Two-Dimensional
        Electron Fluid}
\author{D. Belitz$^{1}$ and T.R. Kirkpatrick$^{2}$}
\affiliation{$^{1}$Department of Physics, Institute of Theoretical Science, and Materials Science Institute, University
                of Oregon, Eugene, OR\\ 
                $^{2}$Institute for Physical Science and Technology, University of Maryland, College Park, MD 20742
                }
\date{\today}
\begin{abstract}
We present a scaling description of a metal-insulator transition in two-dimensional electron
systems that is driven by a vanishing compressibility rather than a vanishing diffusion coefficient.
A small set of basic assumptions leads to a consistent theoretical framework that is compatible
with existing transport and compressibility measurements, and allows to make predictions for
other observables. We also discuss connections between these ideas and other theories of
transitions to an incompressible quantum fluid.
\end{abstract}
\maketitle
The metal-insulator transition (MIT) observed in two-dimensional (2D) electron (and hole) systems remains very incompletely 
understood \cite{Abrahams_Kravchenko_Sarachik_2001, Spivak_et_al_2010, Popovic_2016}. Observationally, 
experiments on high-mobility Si MOSFETs and other 2D systems have shown, increasingly convincingly, that a quantum 
phase transition (QPT) from an insulator to a metal occurs with increasing carrier concentration \cite{Popovic_2016}. 
Theoretically, various explanations have been proposed that include no true transition, percolation, and modifications of 
localization theory \cite{Abrahams_Kravchenko_Sarachik_2001, Spivak_et_al_2010}. Interesting recent 
developments include the realization that the transition in low-mobility/strong-disorder
systems may be in a different universality class than the one in high-mobility/weak-disorder ones \cite{Popovic_2016},
and the observation that the thermopower in the metallic phase shows critical behavior at a critical value 
$n_{\text{c}}$ of the electron density $n$ that coincides, at least in high-mobility/weak-disorder systems, with the critical 
density for the MIT \cite{Mokashi_et_al_2012} .
The latter observation was recently confirmed by a transport study of the insulating phase 
\cite{Li_Sarachik_2016}. Another development are 
measurements of the thermodynamic density susceptibility $\dndmu$ \cite{Dultz_Jiang_2000, Ilani_et_al_2000, Allison_et_al_2006}, 
which is related to the electronic compressibility $K_T$ via $\dndmu = n^2 K_T$. 
These experiments show that $K_T$ vanishes in the insulating phase. Some also find $K_T<0$
in the metallic phase, an observation that we will come back to below.

In this Letter we focus on weakly disordered systems and construct a scaling description of a 2D MIT that is
driven by a vanishing compressibility. We will show that such a description is conceptually consistent and in
agreement with existing experiments, and we will make predictions for the behavior of additional observables. 

The electrical conductivity $\sigma$ is given in terms of the diffusion coefficient $D$ and 
$\dndmu \equiv \chi_n$ by the Einstein relation
\be
\sigma = \chi_n D\ .
\label{eq:1}
\ee
Our description of the 2D MIT is based on two basic assumptions, viz.,

{\it Assumption 1:} The observed MIT is indeed a continuous QPT
characterized by a diverging correlation length $\xi$ with critical exponent $\nu$, $\xi \propto r^{-\nu}$,
where $r = \vert n - n_{\text{c}}\vert/n_{\text{c}}$ is the dimensionless distance from the QPT.

{\it Assumption 2:} The diffusion coefficient $D$ is uncritical, at least for weakly disordered systems. 
This implies that $D$ has its naive scale dimension. Choosing the scale dimension of
$\xi$ as $[\xi] = -1$, we have $[D] = -2 + z_D$, with $z_D$ an appropriate dynamical exponent. We note 
for later reference that there is more than one dynamical exponent. $z_D$ is
related to diffusive physics, hence $z_D = 2$ and $[D] = 0$. This assumption is motivated by the observation
that the transition is not sensitive to weak disorder \cite{Mokashi_et_al_2012, Popovic_2016}. The experimental
evidence for a MIT as interpreted by Assumption 1 then implies, via Eq.~(\ref{eq:1}), that $\chi_n$ vanishes as the 
transition is approached from the metallic side and remains zero in the insulator, i.e., the insulator is an 
incompressible quantum fluid. Assigning a scale dimension $[\mu] = z$ to the chemical potential $\mu$, we have 
$[\chi_n] = 2 - z$. This is a strong scaling assumption in the sense of the term defined in Ref.~\cite{Kirkpatrick_Belitz_2015b}:
It assumes that $[\chi_n]$ simply follows from the fact that $1/\chi_n$ is an energy times a volume and is not affected
by dangerous irrelevant variables. The dynamical exponent $z$ is different from the diffusive $z_D$ above. 
Its value is {\it a priori} unknown and only bounded by $z<2$ (to ensure that $\chi_n$ vanishes). We will discuss plausible 
values of $z$ below.

We note that in standard theories of electron localization, with or without interactions, $[\mu]=d$ in $d$ spatial
dimensions, $\chi_n$ is uncritical, and the MIT is driven by a vanishing diffusion coefficient \cite{Lee_Ramakrishnan_1985, 
Belitz_Kirkpatrick_1994}. In this sense our Assumption 2 is exactly complementary to localization physics. 

The above assumptions immediately lead to homogeneity laws for both $\chi_n$ and $\sigma$,
\bea
\chi_n(r,T) &=& b^{-(2-z)}\,F_{\chi}(r\,b^{1/\nu}, T\,b^z)\ ,
\label{eq:2}\\
\sigma(r,T) &=& b^{-(2-z)}\,F_{\sigma}(r\,b^{1/\nu}, T\,b^z)\ .
\label{eq:3}
\eea
$F_{\chi}$ and $F_{\sigma}$ are scaling functions, $T$ is the temperature, $b>0$ is an arbitrary scale factor, and
we have assumed $[T] = [\mu] = z$. This is the simplest possible dynamical-scaling
assumption; we will discuss generalizations below.

We first focus on $\sigma$. Equation (\ref{eq:3}) implies 
\bea
\sigma(r,T) &=& r^{\nu(2-z)}\,F_{\sigma}(1,T/r^{\nu z})
\nonumber\\
                 &=& T^{(2-z)/z)}\,F_{\sigma}(r/T^{1/\nu z},1)\ .
\label{eq:4}
\eea
Numerous sets of data have been fitted to a scaling law of this form, initially assuming $z=d=2$, which was motivated
by localization physics. Later it was realized that low-$T$ data do not support this assumption. Determining
the exponents precisely from scaling plots has proven difficult; current best estimates from transport measurements
for high-mobility systems are $\nu(2-z) \approx 1 - 1.5$, $(2-z)/z \approx 1.5$, and 
$\nu z \approx 1 - 1.7$ \cite{Popovic_2016, nuz_footnote}. 
Combining these results we obtain $z \approx 0.8$ and $\nu \approx 0.83 - 2.1$.  An independent value for $\nu z\approx 1$ 
is obtained from the thermopower data~\cite{Mokashi_et_al_2012}, which were interpreted by means of a scaling 
theory in Ref.~\cite{Kirkpatrick_Belitz_2013a}. Taken together, these data support
scaling of the conductivity given by Eq.~(\ref{eq:4}) with
\be
z \approx \nu \approx 1
\label{eq:5}
\ee
within substantial error bars. 

The critical behavior of $\dndmu$ affects the screening of the Coulomb interaction. The wave-number and frequency dependent 
dielectric function $\epsilon({\bm q},\omega)$ is \cite{Pines_Nozieres_1989}
\be
\epsilon({\bm q},\omega) = \epsilon_0\,\left[1 + v({\bm q})\,\chi_{\text{sc}}({\bm q},\omega)\right]\ .
\label{eq:6}
\ee
$v({\bm q}) = 2\pi e^2/\epsilon_0\vert{\bm q}\vert$ is the bare 2D Coulomb potential, and
$\epsilon_0$ is the static background dielectric constant. The screened potential is
\be
v_{\text{sc}}({\bm q},\omega) = \frac{v({\bm q})}{\epsilon({\bm q},\omega)/\epsilon_0} = \frac{v({\bm q})}{1 + v({\bm q})\,\chi_{\text{sc}}({\bm q},\omega)}\ .
\label{eq:7}
\ee
$\chi_{\text{sc}}$ 
is the screened density susceptibility, which at small ${\bm q}$ and $\omega$ is diffusive:
\bse
\label{eqs:8}
\be
\chi_{\text{sc}}({\bm q},\omega) = \chi_n\,i D {\bm q}^2/(\omega + i D {\bm q}^2)\ .
\label{eq:8a}
\ee
In particular,
\be
\chi_{\text{sc}}({\bm q}\to 0, \omega=0) = \chi_n\ ,
\label{eq:8b}
\ee
\ese
which expresses the compressibility sum rule. The statically screened potential thus is
\be
v_{\text{sc}}({\bm q},\omega=0) = \frac{2\pi e^2/\epsilon_0}{q + \kappa}\ ,
\label{eq:9}
\ee
with $\kappa = (2\pi e^2/\epsilon_0) \partial n/\partial\mu$ the 2D screening length, which has the same scale
dimension as $\chi_n$, $[\kappa] = 2-z$. If $z=1$, then $\kappa$
scales as an inverse length and the static dielectric function, $\epsilon({\bm q},\omega=0)/\epsilon_0 = 1 + \kappa/\vert{\bm q}\vert$,
scales as a constant.

As the transition is approached from the metallic phase, $\xi \propto r^{-\nu}$ and $1/\kappa \propto r^{-\nu(2-z)}$ both diverge, 
and screening breaks down at the transition. For $z=1$, $1/\kappa \propto \xi$, whereas the dielectric function is uncritical.

The above conclusions all follow from Assumptions 1 and 2. In order to make predictions about the behavior of
other observables we need an additional assumption regarding the dynamics. The simplest possibility is
\smallskip\par\noindent
{\it Assumption 3a:} The strong-scaling assumption of Assumption 2 applies to other observables that have the 
dimensions of a density of states (DOS), including $\dndmu$, the single-particle DOS $N$, and the specific-heat coefficient 
$\gamma$ (but not the spin susceptibility, which is uncritical in our picture of the 2D MIT), and they are all 
governed by the same dynamical exponent $z$. That is, 
apart from the diffusive exponent $z_D=2$ there is only one dynamical exponent $z$ \cite{degenerate_z_footnote, single_z_footnote}. 
Both $N$ and $\gamma$ then obey scaling laws analogous to Eqs.~(\ref{eq:2}, \ref{eq:3}), which leads to
\bse
\label{eqs:10}
\bea
\chi_n(r,T) &=& r^{\nu(2-z)}\,G_{\chi}(T/r^{\nu z})\ ,
\label{eq:10a}\\
\gamma(r,T) &=& r^{\nu(2-z)}\,G_{\gamma}(T/r^{\nu z})\ ,
\label{eq:10b}
\eea
for $\chi_n$ and $\gamma$, and
\be
N(r,\omega,T) = r^{\nu(2-z)}\,G_{N}(\omega/r^{\nu z}, T/r^{\nu z})
\label{eq:10c}
\ee
\ese
for $N$. Here $G_{\chi}(x) = F_{\chi}(1,x)$, and $G_{\gamma}$ and $G_N$ are analogous scaling functions.
$N(r,\omega,T)$ is the single-particle DOS, which can be measured by tunneling, as a function of $r$, $T$, and
the energy distance $\omega$ from the Fermi surface. According to Assumption 3a, the exponents $\nu$ and $z$
are the same as those that govern the scaling of the conductivity, Eqs.~(\ref{eq:4}, \ref{eq:5}).

Also of interest is the thermal expansion coefficient $\alpha_V = -(1/n)(\partial n/\partial T)_p$, and the
Gr{\"u}neisen parameter $\Gamma = \alpha_V/\gamma T$. According to our assumptions, $\alpha_V$ scales the
same as $\chi_n$ and $\gamma$, so at the QPT $\Gamma$ is diverges as
\be
\Gamma(r=0,T\to 0) \propto 1/T\ .
\label{eq:11}
\ee
The difference between this and the result of Ref.~\cite{Zhu_et_al_2003}, which obtained 
$\Gamma \propto 1/T^{1/\nu z}$, can be understood as follows. We rewrite $\alpha_V$ as in Ref.~\cite{Zhu_et_al_2003},
$\alpha_V = -(1/V)(\partial S/\partial p)_T$. $\Gamma$ then scales as the inverse
pressure. Since the compressibility is $K_T = (1/n)(\partial n/\partial p)_T$, $\chi_n \equiv \partial n/\partial\mu$ can be written
$\dndmu = n(\partial n/\partial p)_T$. At the 2D MIT we have $[p] = [\mu] = [T] = z$, 
and we again obtain Eq.~(\ref{eq:11}). At the magnetic QPTs considered in Ref.~\cite{Zhu_et_al_2003}, 
on the other hand, the pressure scales as the control parameter, $[p] = [r] = 1/\nu$, and hence $\Gamma \sim 1/r \sim 1/T^{1/\nu z}$.

In addition to $\nu$ there are six other static exponents that describe the
$r$-dependence of various thermodynamic quantities, see Ref.~\cite{Kirkpatrick_Belitz_2015b}. 
We use the definitions and notations put forward in that reference, some of which deviate by 
necessity from the standard ones used for classical phase transitions. The specific-heat exponent $\bar\alpha$
is obtained from Eq.~(\ref{eq:10b}) as $\bar{\alpha} = -\nu(2-z)$, in agreement with one of the 
hyperscaling relations derived in \cite{Kirkpatrick_Belitz_2015b}. The exponent $\alpha$ is given in terms
of $\nu$ and $z$ by the hyperscaling relation $\alpha = 2 - \nu(2+z)$. We define $\beta$ by
$\chi_n(r,T=0) \propto r^{\beta}$, $\beta = \nu(2-z)$ \cite{beta_footnote}. $\gamma$ then follows
from the Essam-Fisher relation $\gamma = 2 - \alpha - 2\beta$, $\delta$ is given 
by the Widom relation $\delta = 1 + \gamma/\beta$, and $\eta$ 
by the Fisher
relation $\eta = 2 - \gamma/\nu$. With $\nu = z = 1$ this yields $\bar\alpha = \alpha = -1$, $\beta = \gamma = \eta = 1$
and $\delta = 2$. $\alpha$, $\gamma$, $\eta$, and $\delta$ describe the critical behavior of correlation
functions that are non-standard and would be hard to measure. 

Another set of exponents describes the $T$-dependence of observables $r=0$. They are
given by $\omega_T = \omega/\nu z$, where $\omega$ can be $\bar\alpha$, $\alpha$, $\beta$, $\gamma$, or
$\nu$. With $\nu z = 1$ from the thermopower experiment \cite{Mokashi_et_al_2012} we have $\omega_T = \omega$. 

The exponents $s$ and $s_T$ that define the dependence of the
electrical conductivity $\sigma$ on $r$ and $T$, respectively, are $s = \nu z s_T = \nu(2-z)$, see Eq.~(\ref{eq:4}). Note that $\sigma$ does
not obey the strong-scaling assumption that is valid in localization physics and leads to Wegner scaling, i.e., $s = \nu(d-2)$,
see Refs.~\cite{Lee_Ramakrishnan_1985, Belitz_Kirkpatrick_1994, Kirkpatrick_Belitz_2015b}. Rather, $D$ and $\chi_n$
in Eq.~(\ref{eq:1}) are governed by separate strong-scaling assumptions. 

So far we have assumed that there is only one dynamical exponent $z$. 
This is possible, and the experimental data to date are, within the error bars, consistent with this notion. However,
in general QPTs are characterized by more than one dynamical critical 
exponent \cite{Belitz_Kirkpatrick_1994, von_Loehneysen_et_al_2007, Kirkpatrick_Belitz_2015b, multiple_z_footnote}.
For instance, the theory of a continuous 2D Mott transition in Ref.~\cite{Senthil_2008} contains two dynamical
exponents, $z_1 = 2$ and $z_2 = 1$. 
The latter governs bosonic degrees of freedom
that arise from decoupling the fermions, and the former is due to a gauge field associated with spinon 
degrees of freedom. Another example of a QPT with two dynamical exponents is the quantum
ferromagnetic transition \cite{Kirkpatrick_Belitz_2015b, Brando_et_al_2016a}. This leads to an
\smallskip\par\noindent
{\it Assumption 3b:} The QPT is characterized by dynamical exponents $z_n$ ($n = 1,2,\ldots$) such
that $z_1 > z_2 > \ldots$ The specific heat coefficient $\gamma$, which is the energy-energy susceptibility,
will then be governed by the largest $z$. That is, $\gamma$ obeys Eq.~(\ref{eq:10b}) with $z = z_1$,
\be
\gamma(r,T) = r^{\nu(2-z_1)}\,G_{\gamma}(T/r^{\nu z_1})\ .
\label{eq:12}
\ee
For other observables no general statements can be made. Naive scaling suggests that
$z_1$ governs the dynamics of {\em all} observables, but in general this is not true since some observables
do no couple to the leading dynamical processes or, in an alternative interpretation, a subleading time scale
acts as a dangerous irrelevant variable \cite{Belitz_Kirkpatrick_1994}. For
instance, in quantum ferromagnets the order parameter (OP) and its derivatives are governed by a subleading
time scale \cite{Kirkpatrick_Belitz_2015b, Brando_et_al_2016a}, and so is the compressibility in 
Ref.~\onlinecite{Senthil_2008}. In both of these examples the specific heat is governed
by the largest $z$, as expected.

We conclude with various discussion remarks, organized into four groups.
\smallskip\par\noindent
(1) Nature of the transition:
\smallskip\par
(i) Our most basic assumption is that the observed 2D MIT is a quantum phase transition from a
metal to an insulator that is an incompressible quantum fluid, which is one of the hallmarks
of a transition of Mott type \cite{Imada_Fujimori_Tokura_1998}. Another assumption is that the
transition is continuous. 
For our scaling theory we {\em assume}, based on experimental evidence, that $\dndmu$ vanishes 
at the transition; explicit mechanisms for a mechanisms for a vanishing $\dndmu$ have been proposed
in Refs.~\cite{Si_Varma_1998} and \cite{Senthil_2008}.

(ii) We do not assume that there is an OP for this transition, nor do we specify what it is if there is one. All we
assume is scaling, which is much more general. An example of a QPT with no OP, at least not
a conventional one, is the Anderson transition of noninteracting electrons \cite{Evers_Mirlin_2008}. For the
Mott-type transition discussed in Ref.~\cite{Senthil_2008} it is not obvious whether or not an OP
exists, yet scaling works and the results agree with the general scaling description of QPTs developed in
Ref.~\cite{Kirkpatrick_Belitz_2015b}. If an OP description of the 2D MIT is possible, then
$\dndmu$ is an obvious candidate \cite{beta_footnote}. Another candidate is the single-particle DOS, which serves as the
OP for the Anderson-Mott transition \cite{Kirkpatrick_Belitz_1994, Belitz_Kirkpatrick_1995}. Interestingly, 
in this OP theory for the Anderson-Mott transition $\dndmu$ is also
critical and scales the same way as the DOS. Our premise is that the same is true for the 2D MIT.

(iii) An important consequence of our assumptions is that the density response remains diffusive as one approaches
the critical point from the metallic side, only the prefactor of the diffusion pole vanishes.

(iv) Spin does not enter our discussion, and we therefore expect an in-plane magnetic field, even one
strong enough to completely polarize the electrons, to have no qualitative effect on the transition.
This is in agreement with the experimental observations \cite{Li_Sarachik_2016, Popovic_2016}.

(v) An interesting question is the nature of the metallic phase. If it is a Fermi liquid (FL), then the transition
is from a compressible FL to an incompressible quantum fluid, along the lines of
Refs.~\cite{Kirkpatrick_Belitz_2013a, Kirkpatrick_Belitz_2012}. However, this requires an unknown mechanism by 
which strong correlations suppress localization in 2D. 
A candidate for such a mechanism is discussed in point (3) (ii) below.

\medskip\par\noindent
(2) Exponent values:
\smallskip\par
(i) The experiments are consistent with $z=1$, see Eq.~(\ref{eq:5}). If the metallic phase is a FL,
then there will be at least one $z$ that is equal to 1, since FLs are characterized
by ballistic soft modes whose frequency scales as the wave number \cite{Belitz_Kirkpatrick_2012,
Belitz_Kirkpatrick_2014, z=1_footnote}. However, the metallic phase may well be a ``strange metal'' rather than a
FL: Tuning the electron-electron interaction to zero will turn the metal into an insulator due
to weak-localization effects, so it is unlikely to be a FL. The observations of a negative
compressibility are also indicative of a non-standard metal. 

(ii) If metallic soft modes related to those in a FL are indeed the source of $z=1$, then
disorder will change this exponent to $z=2$, which is characteristic of the diffusive soft modes in
a disordered metal. This suggests that the scale dimension of $\dndmu$ will be smaller
($2-z = 0$ plus a likely small anomalous dimension) in disordered samples than in clean ones.
However, even in the disordered case there are likely still soft modes with $z=1$, e.g., the
bosonic excitations in Ref.~\cite{Senthil_2008}. It is thus possible that there are various degenerate
dynamical exponents in the clean case whose degeneracy is lifted by disorder. 

(iii) Irrespective of the origin of the observed $z=1$, the thermopower experiment of Ref.~\cite{Mokashi_et_al_2012}
implies $\nu = 1$, provided there either is only one dynamical exponent, or the thermopower depends on a $z$
that is equal to 1. The remaining static exponents then follow from various (hyper)scaling relations, as
discussed in the main text. Some of these exponents values are unusual, e.g., $\delta = 2$ is unusually small.
This reflects long-range correlations that are effectively built into our scaling scenario.

(iv) A generic mechanism for generating a $z>1$, which will then determine the critical dynamics of the
specific heat, see Eq.~(\ref{eq:12}), is Landau damping, which occurs if bosonic degrees of freedom couple
to fermionic ones. This mechanism is very general and thus expected to be widespread; explicit examples
include the OP fluctuations in quantum ferromagnets, which have a $z=d$ due
to Landau damping \cite{Kirkpatrick_Belitz_2015b, Brando_et_al_2016a}, and the gauge fluctuations
in Ref.~\cite{Senthil_2008}, which have a $z=2$ by the same mechanism.

\medskip\par\noindent
(3) The role of disorder:
\smallskip\par
(i) Our scaling scenario is motivated by the observed 2D MIT in high-mobility samples, where 
experiments indicate that the transition is not qualitatively affected by disorder \cite{Mokashi_et_al_2012,
Popovic_2016}.We point out, however, that these samples are not weakly disordered in any
conventional sense. While the mean-free path is long due to the high mobility, the Fermi wavelength
$1/\kF$ is equally long due to the low Fermi temperature (typically, a few K \cite{Lin_Popovic_2015}).
As a result, $\kF\ell$ is not large compared to 1 even in high-mobility samples, which raises the
question of what suppresses localization effects. This issue is not understood; a possible resolution
is discussed in the following discussion point.

(ii) An important observation is the negative compressibility in the metallic phase
\cite{Kravchenko_et_al_1990, Dultz_Jiang_2000, Allison_et_al_2006}, which is also found in quantum Hall 
systems \cite{Eisenstein_Pfeiffer_West_1992}, and in a quasi-3D metal
\cite{He_et_al_2015}, and the physical meaning of which is not understood. $K_T<0$
makes the system mechanically unstable and most likely indicates a spontaneous rearrangement into
a spatially inhomogeneous state \cite{Eisenstein_1992}. This may be an important ingredient for
stabilizing a metallic phase in 2D system, as it will introduce a new length scale that cuts off the
weak-localization singularities. As a result, the effects of quenched disorder may be much weaker than
in other 2D systems.

\medskip\par\noindent
(4) Suggestions for experiments:
\smallskip\par
(i) It would be very interesting to measure the tunneling DOS of the same samples that show critical
behavior of the electrical conductivity and the thermopower. It follows from Assumption 3a that the
DOS will obey a homogeneity law, Eq.~(\ref{eq:10c}), with the same exponents $\nu$ and $z$ as
$\dndmu$ and hence $\sigma$. 

(ii) Of equal interest would be measurements of the specific heat of the same samples. In the presence 
of multiple dynamical exponents the specific heat 
depends on the largest one, Eq.~(\ref{eq:12}), so a specific-heat coefficient that obeys Eq.~(\ref{eq:12}) with $z_1>1$ 
would be a clear indication of multiple dynamical exponents. 

\acknowledgments
We  thank Myriam Sarachik for discussions. This work was supported by the NSF under Grants No. 
DMR-1401410 and No. DMR-1401449.  Part of this work was performed at the Aspen Center for Physics 
and supported by the NSF under Grant No. PHY-1066293.


\begin{thebibliography}{39}
\bibitem[{deg()}]{degenerate_z_footnote}
\bibinfo{note}{This includes the possibility that there are several degenerate
  dynamical exponents, but for scaling purposes there is only one.}
\bibitem[{bet()}]{beta_footnote}
\bibinfo{note}{If $\chi_n$ can be considered the order parameter for the
  transition, then the such defined $\beta$ has the standard interpretation as
  the order-parameter exponent. Otherwise the exponent relations from
  Ref.~\cite{Kirkpatrick_Belitz_2015b} still hold, but $\beta$, $\gamma$,
  $\eta$, and $\delta$ do not have their standard meaning.}
\end{thebibliography}

\end{document}